\documentclass[journal]{IEEEtran}
\usepackage{graphicx}
\usepackage{amsmath}
\usepackage[utf8]{inputenc}
\usepackage[]{xcolor}
\usepackage[T1]{fontenc}
\usepackage[none]{hyphenat}
\usepackage{multirow}

\usepackage{cite}

\ifCLASSINFOpdf
\else
\fi

\begin{document}
%
\title{Design and Validation of Metasurface Transmitarrays for Time-Reversal Microwave Hyperthermia in Deep Brain Tumors}
%
%
\author{Alireza Rahmani,
        Mohammad Javad Hajiahmadi,
        and~Reza Faraji-Dana,~\IEEEmembership{Senior Member,~IEEE}%
\thanks{Manuscript received Month DD, YYYY; revised Month DD, YYYY. }%
\thanks{M. J. Hajiahmadi is with  Department of Electrical Engineering, Iran University of Science and Technology, Tehran, Iran (e-mail: hajiahmadi@iust.ac.ir).}%
\thanks{A. Rahmani and R. Faraji-Dana are with  School of Electrical and Computer Engineering, University of Tehran, Tehran, Iran}%
}

\maketitle
\begin{abstract}
This paper presents a metasurface-based microwave hyperthermia system for deep-seated brain tumors, designed using a systematic time-reversal approach. The excitation phase of the metasurface is derived from time-reversal of a point source located at the tumor, enabling precise manipulation of the transmitted wavefront. A quad-layer transmissive metasurface, formed by square metallic patches, is engineered to realize the required phase-transfer function. Numerical simulations on realistic and spherical head models demonstrate effective electromagnetic focusing, achieving a normalized power loss density figure of merit $\mathrm{FoM_Q}=0.769$, indicating minimal undesired heating of healthy tissue. An electrothermal head phantom with dielectric constant 43 and conductivity 0.9 was fabricated to experimentally validate the system at 1.8 GHz. Experimental results show a temperature increase of approximately 5°C at the tumor site after 20 minutes of exposure, while healthy regions remain within 2°C of baseline. To the authors’ knowledge, this is the first experimental demonstration of metasurface-based time-reversal focusing for deep brain hyperthermia, opening a pathway for noninvasive, highly localized treatment.

\end{abstract}

\begin{IEEEkeywords}
Brain tumor, microwave hyperthermia, metasurface, time-reversal, transmit array.
\end{IEEEkeywords}

%
\IEEEpeerreviewmaketitle

\section{Introduction}
%
%
%
%
\IEEEPARstart{M}{icrowave} Microwave hyperthermia represents a non-invasive therapeutic modality, typically employing electromagnetic waves within the frequency range of 915 MHz to 2.45 GHz. This frequency band facilitates significant penetration of electromagnetic fields into biological tissues, coupled with an elevated rate of heat absorption by water molecules \cite{hyperthermia1,hyperthermia2}. The treatment involves the application of an external heat source to selectively elevate the temperature of the tumor region to a range of 40 to 45 degrees Celsius. This thermal effect is primarily driven by the absorption of microwave energy by water molecules present within the tissue~\cite{hyperthermia3,hyperthermia4,hyperthermia5}. 

Microwave hyperthermia applicators are conventionally classified into two distinct categories based on their design and operational characteristics. The predominant systems employed in microwave hyperthermia utilize phased array antennas to achieve targeted thermal delivery \cite{array1,array2,array3,array4,array5,array6,array7,array8,array9}. These systems leverage a multitude of antennas to precisely focus electromagnetic fields at specific anatomical locations. The efficacy of phased array-based hyperthermia applicators has been enhanced through the implementation of time-reversal techniques, which improve the precision of energy deposition \cite{tr1,tr2,tr3}. Despite their capability for narrow and accurate focusing, these systems necessitate complex and extensive antenna arrays, alongside intricate antenna feeding mechanisms, which pose significant design and operational challenges \cite{ms_r_1,ms_r_2,ms_r_3,ms_r_4}. 

An alternative approach involves the use of metamaterial-based applicators, whose suitability for hyperthermia applications has been explored \cite{article_meta1,article_meta2}. Specifically, applicators incorporating left-handed lenses, derived from metamaterial principles, have been proposed \cite{article_meta3,article_meta4}. Additionally, metasurface-based microwave hyperthermia systems have emerged as a promising avenue \cite{9912308}, driven by the growing interest in metasurface applications within bioelectromagnetics \cite{metasurface_bio1,metasurface_bio2}. The manipulation of the incident wave in the transmissive metasurfaces can be obtained either through the change in the physical properties of the unit cell, for example, patch length and number of layers \cite{unit_static1,unit_static2,unit_static3,unit_static4,unit_static5} or through a reconfigurable element such as Varactor and PIN diodes \cite{u1,u2,u3,u4,u5,u6}. However, a critical limitation of metamaterial-based systems is the paucity of experimental validation, which hinders their clinical translation and practical implementation.

In this study, we present a transmit array antenna designed to function as a microwave hyperthermia applicator for the treatment of deep-seated brain tumors, leveraging its high-resolution, controllable wavefront shaping capability. Numerical simulations demonstrate that the proposed four-layer unit-cell structure enables precise manipulation of the transmission phase by adjusting the unit-cell geometry sizes on the metasurface plane, thereby accommodating individualized tumor locations for each patient. The efficacy of this applicator for brain hyperthermia is evaluated through comprehensive simulations. Moreover, experimental validations conducted on a synthesized thermo-electric phantom demonstrate that the applicator achieves a targeted temperature increase of approximately 2°C at the tumor site, while ensuring that surrounding healthy tissues remain within safe thermal thresholds.

In Section II, we first introduce the statemetn of the problem and setup of the proposed microwave hyperthermia system. In Section III, the principles of the time-reversal technique and metasurface design are presented, and the proposed transmissive unit cell is introduced. In Section IV, the numerical investigation of the proposed metasurface is shown later. Section V contains the modifications that are applied to the proposed microwave hyperthermia system for fabrication, and the numerical results are presented in the modified system. Fabrication of the metasurface and the corresponding head phantom model and the altered numerical results are given in Section VI. Section VII covers the experimental results and discussions on the realized system. Finally, in Section VIII, we draw conclusions based on the proposed system.


\section{Statement of the Proposed Transmit Array Microwave Hyperthermia Method}

\begin{figure}[h!]
\includegraphics[width=3in]{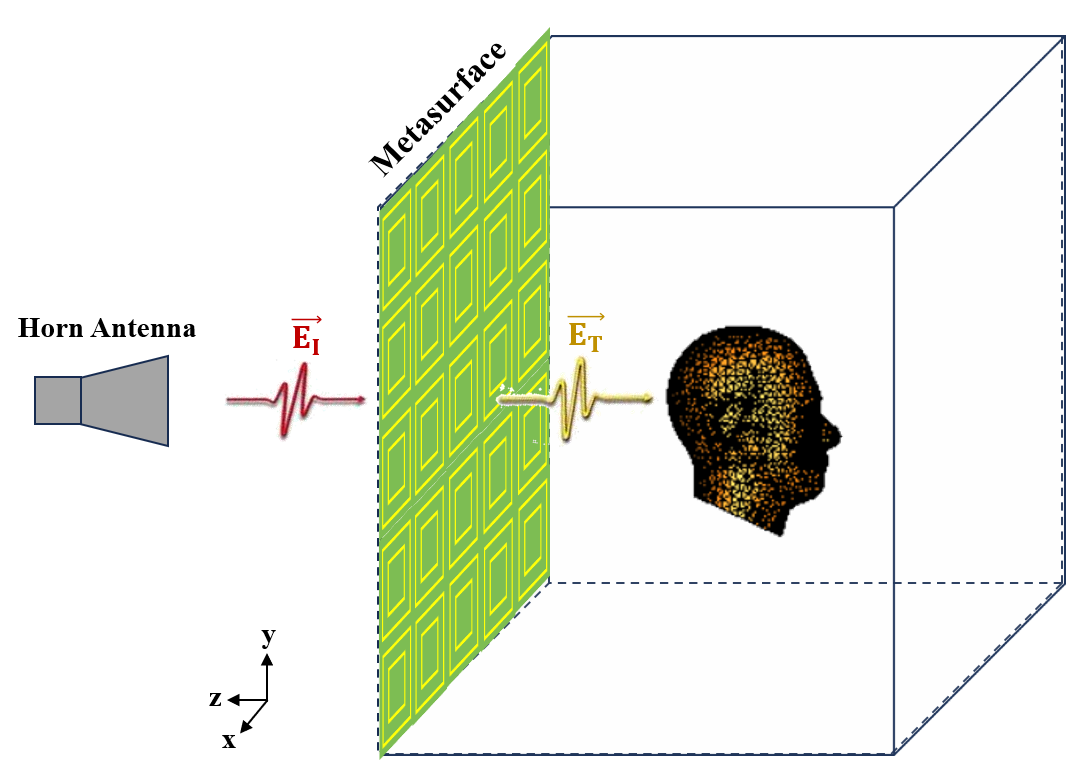}
\centering
\caption{Proposed metasurface-based system of microwave hyperthermia for brain tumors.}
\label{setup_schem}
\end{figure}

Figure~\ref{setup_schem} illustrates a novel transmit-array time-reversal focusing setup designed for microwave hyperthermia therapy targeting deep-seated brain tumors. The configuration consists of four key elements: (1) a transmitting antenna, (2) a transmit-array layer in the form of a transmissive metasurface, (3) a semi-cavity enclosure where the metasurface forms one side, three sides and the top are metallic panels, and the bottom is open to accommodate the patient’s head. The antenna, located in the far-field region and oriented perpendicularly to the metasurface, projects electromagnetic fields uniformly onto the array. The engineered metasurface then redirects these waves into the enclosure, where the fields are focused precisely at the tumor site using the time-reversal principle, as detailed in Section~\ref{subsection_tr}. The surrounding metallic walls, acting as Perfect Electric Conductors (PECs), further enhance the focusing efficiency~\cite{9330350}.

 The time-reversal cavity is cubic, with dimensions of $5\lambda$ at a frequency of 2~GHz. Four PEC walls serve as scatterers to improve focusing~\cite{9330350}, one face is left open for patient access, and the opposite face incorporates the metasurface for wave manipulation. The head model consists of a homogeneous brain surrounded by a thin shell, with permittivity and conductivity values listed in Table~I. The brain dimensions are approximately 17~cm, 29~cm, and 24~cm along the $x$, $y$, and $z$ axes, respectively, representing an average adult brain. A spherical tumor model with a radius of 5~mm is embedded at a random position $(x,y,z) = (10,0,-20)$~mm inside the brain, with dielectric properties matched to brain tissue. The metasurface is excited by a plane wave, and the transmit array design manipulates the incident wavefront to achieve a narrow focus at the tumor center, with the systematic design considerations described in subsequent sections.

\begin{table}[h!]
\centering
\label{tab0}
\caption{Head model properties \cite{head_p}}
\begin{tabular}{ccc} 
\hline
material & $\epsilon_r$ & $\sigma [\frac{S}{m}]$ \\
\hline
\hline
Brain & 43 & 1.32 \\
\hline
Shell & 4.7 & 0.41 \\
\hline
\end{tabular}
\end{table}

\section{Metasurface design procedure}
\label{section_design}
 In this section, we present the underlying principles of the proposed metasurface-based time-reversal focusing system. We begin with a brief overview of the time-reversal technique, followed by the fundamental concepts of the metasurface design. The characteristics of the proposed unit cell are then obtained through simulation. Finally, quantitative metrics are provided to evaluate the performance of the system.  

\subsection{Time-Reversal Procedure}
\label{subsection_tr}
The time-reversal process begins by placing a point source, approximated as a small dipole, at the tumor location. The resulting electromagnetic fields are then recorded on the metasurface aperture through full-wave simulation. Using Eq.~(\ref{E_tr}, \ref{H_tr}), the transverse electric and magnetic fields are calculated. These fields form the basis for the metasurface design, as their accurate reconstruction on the metasurface enables the electromagnetic energy to be refocused precisely at the tumor site.  

\begin{equation} \label{E_tr}
\mathbf{E}_{\mathbf{rev}}\ \left(x,y,z\right)=\mathbf{E}_{\mathbf{rec}}^\ast\ \left(x,y,z\right)
\end{equation}
\begin{equation} \label{H_tr}
\mathbf{H}_{\mathbf{rev}}\ \left(x,y,z\right)=-\ \mathbf{H}_{\mathbf{rec}}^\ast\ \left(x,y,z\right)
\end{equation}
where $\mathrm{E_{rev}}$ and $\mathrm{H_{rev}}$ denote the reconstructed time-reversed fields, and $\mathrm{E_{rec}}$ and $\mathrm{H_{rec}}$ represent the recorded electromagnetic fields on the metasurface aperture.

\subsection{Metasurface Design}
The objective of the metasurface is to match the transmitted phase profile with the phase of $\mathrm{E_{rev}}$ obtained from the time-reversal procedure. Assuming plane-wave excitation with an electric field polarized along the $x$-direction, the corresponding electromagnetic fields are expressed in Eq.~(\ref{plane_wave}).  

\begin{equation} \label{plane_wave}
\begin{split}
\mathbf{E_i}=\ E_0\ \hat{x} \\
\mathbf{H_i}=\ \frac{E_0}{\eta}\ \hat{y}
\end{split}
\end{equation}

The transmission scattering parameters of the metasurface are derived from Eq.~(\ref{s21}). For a plane-wave excitation, the incident phase across the surface is uniform, implying that the phase of $\mathrm{S_{21}}$ corresponds to the transmitted phase. Consequently, the metasurface unit cells must be designed such that their transmission phase equals the phase of the time-reversed electric field $\mathrm{E_{rev}}$.  

\begin{equation} \label{s21}
\begin{split}
S_{21}^{xx}=\ T_{xx}=\ \frac{E_t}{E_i} \\ 
\angle S_{21}=\angle E_t-\angle E_i
\end{split}
\end{equation}

\begin{figure}[h!]
\includegraphics[width=3in]{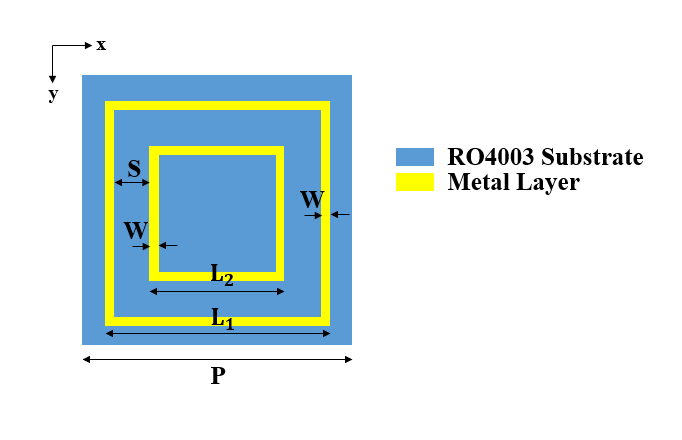}
\centering
\caption{Proposed metasurface schematic.}
\label{metasurface_schem}
\end{figure}

The schematic of the proposed metasurface is shown in Fig.~\ref{metasurface_schem}. The unit cell consists of square metallic patches printed on an RO4003C substrate, with its physical parameters listed in Table~II. The unit-cell period is 30~mm, approximately one-fifth of the free-space wavelength at 2~GHz. Increasing the unit-cell size reduces the focusing accuracy, whereas reducing it increases fabrication and design complexity; thus, a moderate value was selected based on simulation results.  

\begin{table}[h!]
\centering
\label{tab1}
\caption{Proposed metasurface unit cell physical parameters}
\begin{tabular}{cc} 
\hline
Unit cell parameter & Value \\
\hline
\hline
P & 30 [mm] \\
\hline
Substrate thickness & 3 [mm] \\
\hline
Substrate air gap & 13.5 [mm] \\
\hline
W & 1 [mm] \\
\hline
S & $0.2*\mathrm{L_1}$ \\
\hline
\end{tabular}
\end{table}

\begin{figure}[h!]
\includegraphics[width=3in]{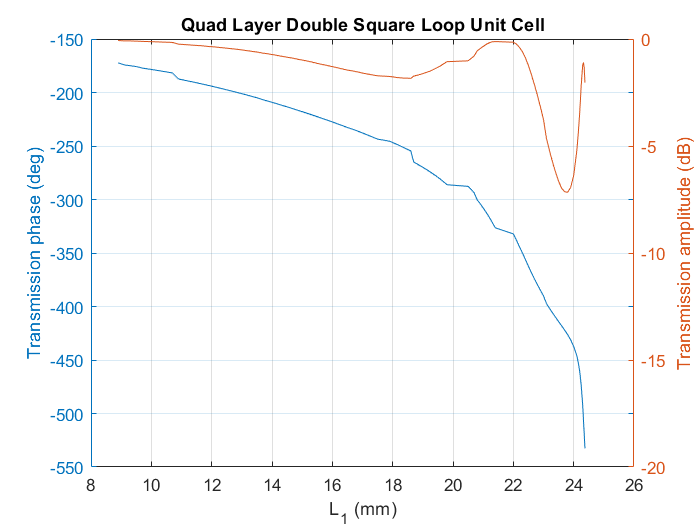}
\centering
\caption{Proposed metasurface transmission phase and amplitude.}
\label{metasurface_transmission}
\end{figure}

The unit-cell performance was characterized using full-wave simulations in CST Microwave Studio with periodic boundary condition. The simulated transmission phase and amplitude responses are shown in Fig.~\ref{metasurface_transmission}. The metasurface achieves a full 360° phase coverage for $\mathrm{L_1}$ ranging from 8.9~mm to 24.4~mm, enabling precise manipulation of the incident wavefront. This capability is achieved through the use of a four-layer metasurface configuration.  

\subsection{Figure of Merit}
The primary objective of the microwave hyperthermia system is to maximize the temperature rise at the tumor site while minimizing undesired heating in surrounding healthy tissues. To quantify this trade-off, we define a Figure of Merit (FoM), expressed in Eq.~(\ref{eq_fom}). Here, $\mathrm{Q_{max}^{healthy}}$ denotes the maximum power loss density in healthy brain tissue, and $\mathrm{Q_{avg}^{tumor}}$ represents the average power loss density within the tumor region. The goal is to minimize this parameter to ensure localized heating at the tumor.  

\begin{equation} \label{eq_fom}
 {FoM}_Q=\ \frac{Q_{max}^{healthy}}{Q_{avg}^{tumor}}
\end{equation}

The power loss density $\mathbf{Q}(\mathbf{r})$ at an arbitrary location $\mathbf{r}$ is given by Eq.~(\ref{Q}), where $\sigma$ denotes the electrical conductivity and $\mathbf{E}$ is the electric field intensity.  

\begin{equation} \label{Q}
\mathbf Q\left(\mathbf{r}\right)=\frac{1}{2}\sigma\left(\mathbf{r}\right)\left|\mathbf{E}\left(\mathbf{r}\right)\right|^2
\end{equation}

\section{Numerical Simulation Results}
\begin{figure}[h!]
\includegraphics[width=3in]{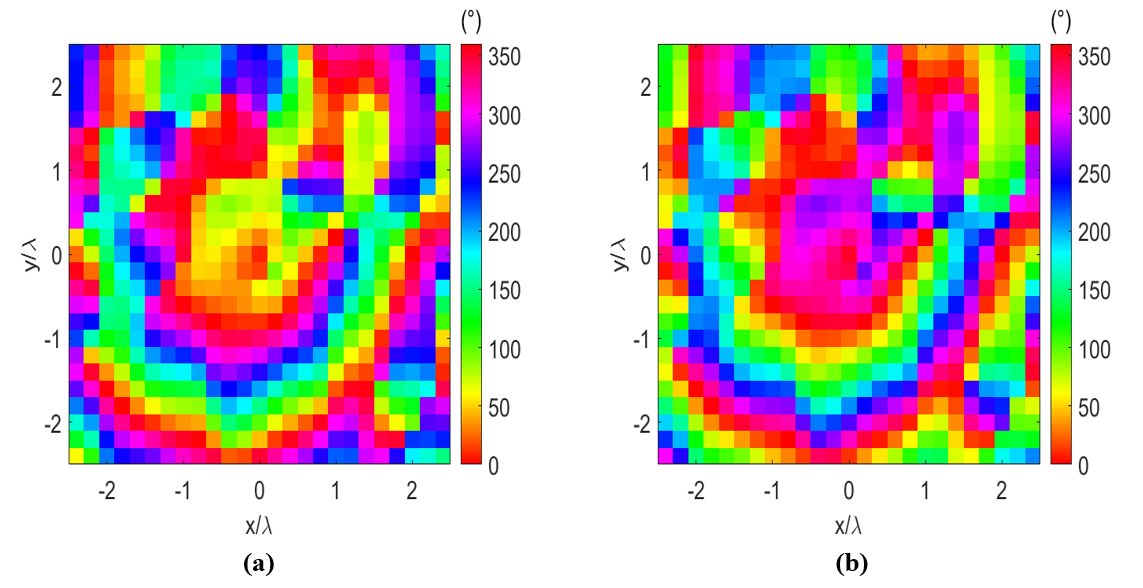}
\centering
\caption{Electric-field phase on the metasurface with PEC walls: (a) received phase, (b) conjugated phase.}
\label{tr_with}
\end{figure}

The systematic design procedure for the proposed metasurface, based on the time-reversal approach, is summarized as follows. First, a small electric dipole is placed at the tumor location to extract the phase of the electric and magnetic fields on the metasurface aperture. In this work, an $x$-oriented dipole is used because the presence of two PEC walls along the $x$-direction reduces the FoM parameter. After phase conjugation, the required transmission phase of the metasurface is obtained. Using MATLAB, the appropriate patch size is selected from the phase-transfer function of the proposed metasurface. Figure~\ref{tr_with} illustrates the received and conjugated electric-field phases on the metasurface.  

\begin{figure}[h!]
\includegraphics[width=3.5in,height=3.25in]{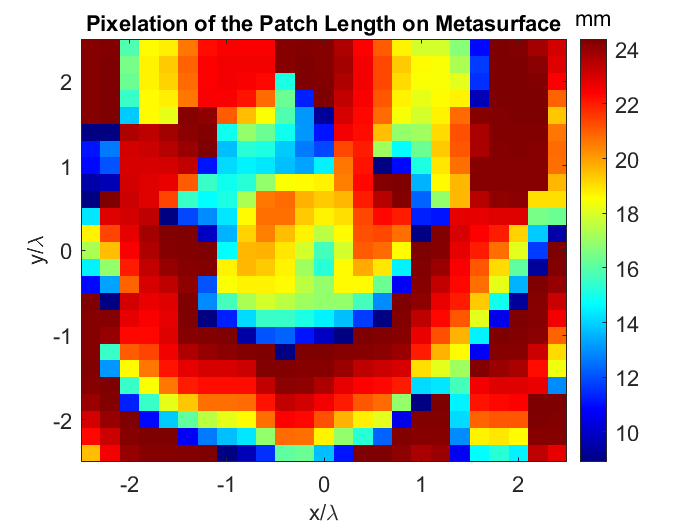}
\centering
\caption{Distribution of $\mathrm{L_1}$ values across metasurface pixels for the PEC-wall configuration.}
\label{length_with}
\end{figure}

\begin{figure}[h!]
\centering
\begin{tabular}{c}
\includegraphics[width=0.48\textwidth, height=1.2in]{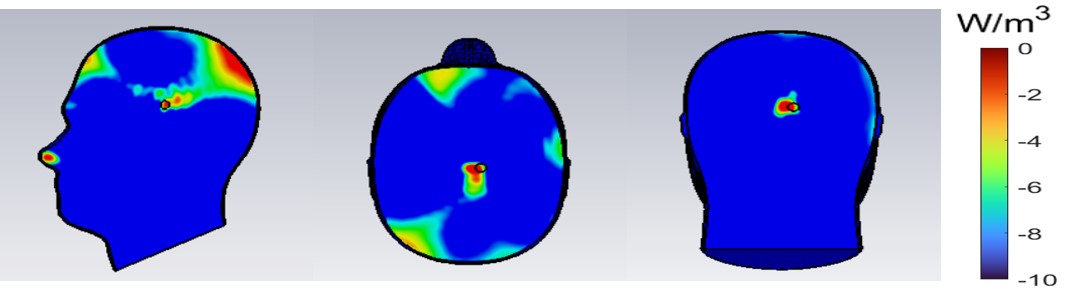} \\
\footnotesize(a) Metasurface implementation ($FoM_Q=1.762$) \\
\includegraphics[width=0.48\textwidth, height=1.2in]{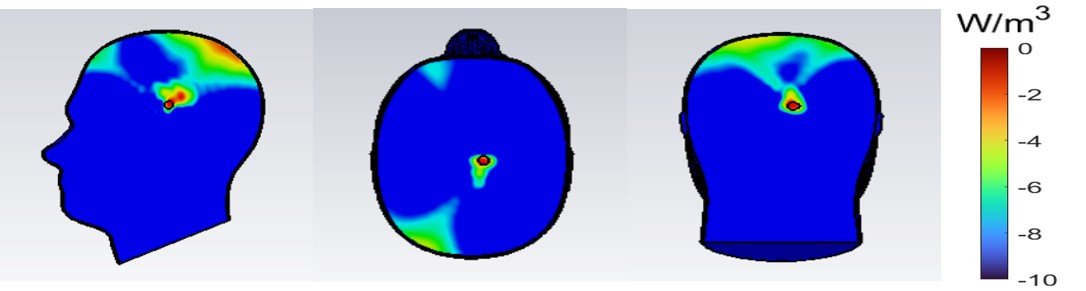} \\
\footnotesize(b) Ideal time-reversal phase sheet ($FoM_Q=1.001$)\\
\end{tabular}
\caption{Normalized power-loss density distribution in different cross-sections (sagittal, coronal, and axial) through the tumor center for the PEC-wall configuration: (a) metasurface implementation, (b) ideal time-reversal phase sheet.}
\label{q_with}
\end{figure}

Since the conjugated electric-field phase matches the phase of $\mathrm{S_{21}}$, the parameter $\mathrm{L_1}$ for each unit cell is obtained using the phase-transfer function shown in Fig.~\ref{metasurface_transmission}. The resulting $\mathrm{L_1}$ distribution across the metasurface pixels is presented in Fig.~\ref{length_with}. A full-wave CST simulation with plane-wave excitation is then performed, and the normalized power loss density is shown in Fig.~\ref{q_with}. These results demonstrate that the metasurface effectively manipulates the wavefront and focuses electromagnetic energy at the tumor location.  

Table~III compares the numerical values of $\mathrm{FoM_Q}$ for two cases: (i) the ideal case, where the exact time-reversal phase is directly imposed on the aperture, and (ii) the metasurface case, where performance is determined by the discrete phase-transfer function under plane-wave excitation. The observed difference arises from quantization errors during the selection of patch lengths, as the phase-transfer function is discretized with a step size of 50~$\mu$m.  

\begin{table}[h!]
\centering
\label{tab2}
\caption{Comparison of $\mathrm{FoM_Q}$ between metasurface implementation and ideal time-reversal phase in the PEC-wall configuration.}
\scriptsize
\begin{tabular}{ccc} 
\hline
Simulation Setup & Metasurface simulation & Ideal time-reversal phase \\
\hline
\hline
With PEC walls & 1.762 & 1.001 \\
\hline
\end{tabular}
\end{table}

\section{Experimental Validation}

After confirming the effectiveness of the proposed metasurface-based microwave brain hyperthermia system through numerical simulations, we proceeded with experimental validation. To achieve this, modifications to the design were implemented to ensure feasibility, followed by the realization of the physical prototype, phantom model, and final measurements. The overall procedure is organized into three subsections: (i) experimental setup modifications, (ii) realization of the system components, and (iii) validation of the complete system.

\subsection{Experimental Setup Modifications}
\begin{figure}[h!]
\includegraphics[width=3in]{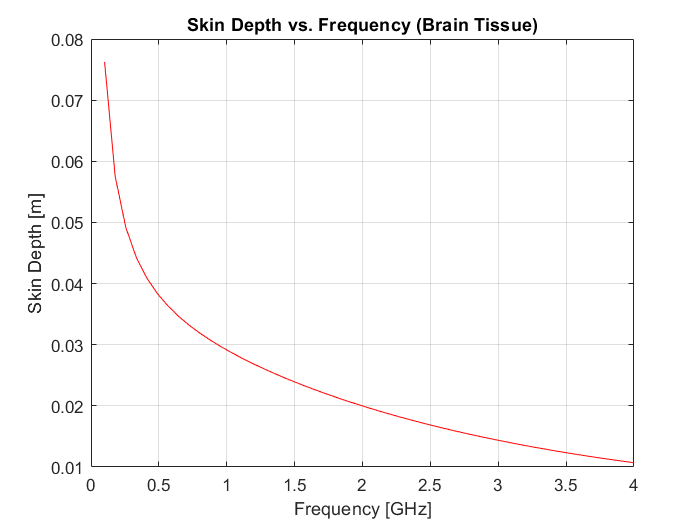}
\centering
\caption{Dispersion diagram of skin depth for brain tissue at microwave frequencies.}
\label{depth}
\end{figure}

Since the experimental tests were conducted in the Type Approval Laboratory of the University of Tehran, the operating frequency was shifted to 1.8 GHz to match the available high-power signal generator. The skin depth at this frequency is given by Eq.~(\ref{skin_depth_eq}), where $\mathrm{\omega}$, $\mathrm{\mu}$, $\mathrm{\sigma}$, and $\mathrm{\varepsilon}$ are the angular frequency, magnetic permeability, electrical conductivity, and permittivity, respectively. Fig.~\ref{depth} shows the dispersion diagram for brain tissue \cite{ITIS_Foundation2024-kx}, indicating that the skin depth at 1.8 GHz is approximately 22 mm.

\begin{equation} \label{skin_depth_eq}
\delta = \sqrt{\frac{2}{\omega\mu\sigma}} 
\sqrt{\sqrt{1+{\left(\frac{\omega\varepsilon}{\sigma}\right)}^2}+\frac{\omega\varepsilon}{\sigma}}    
\end{equation}

The substrate was changed to FR-4 with a thickness of 1.6 mm due to its availability and low cost, while other unit-cell parameters remained unchanged (Table II).  
To validate the system experimentally, a spherical head phantom model with a 140 mm diameter was designed to approximate the size of an adult head.

\subsubsection{Modified Unit-Cell Analysis}
\begin{figure}[h!]
\begin{tabular}{c}
\includegraphics[width=3in]{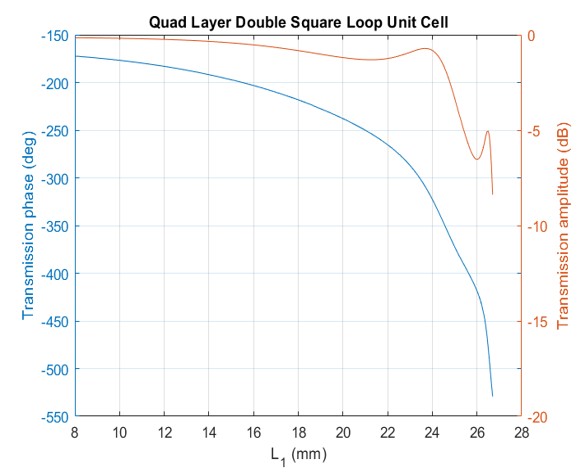} \\
\footnotesize(a) Transmission phase and amplitude of the modified metasurface unit cell \\
\includegraphics[width=3.25in,height=2.75in]{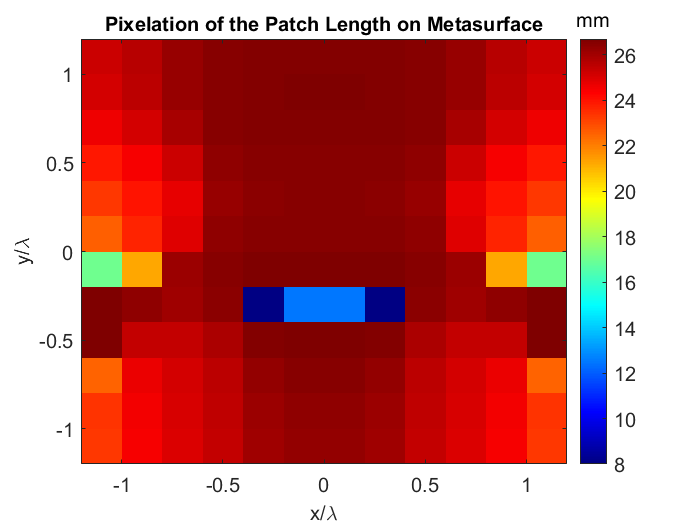} \\
\footnotesize(b) Distribution of $\mathrm{L_1}$ across metasurface pixels\\
\end{tabular}
\caption{Simulation setup modifications using the spherical head model.}
\label{sphere_change}
\end{figure}

The transmission phase and amplitude of the modified unit cell, adapted for ease of fabrication, are shown in Fig.~\ref{sphere_change}(a). After placing an x-oriented dipole at the tumor center $(x, y, z) = (0, 2.5, -10)$ mm, the phase of the transverse electric field was extracted. Since this phase was identical to $\mathrm{S_{21}}$, the nearest parameter $\mathrm{L_1}$ in each unit cell was obtained, as shown in Fig.~\ref{sphere_change}(b).

\subsubsection{Numerical Simulations with the Spherical Head Model}
\begin{figure}[h!]
\centering
\begin{tabular}{c}
\includegraphics[width=0.48\textwidth, height=1.2in]{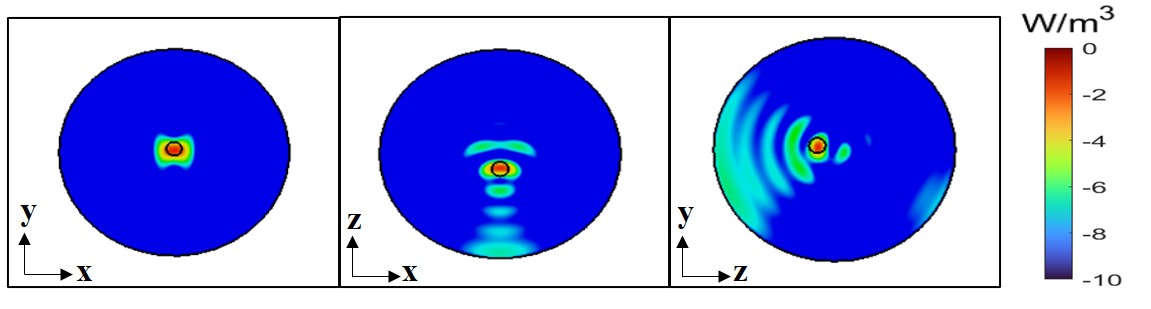} \\
\footnotesize(a) Metasurface simulation ($FoM_Q=0.769$) \\
\includegraphics[width=0.48\textwidth, height=1.2in]{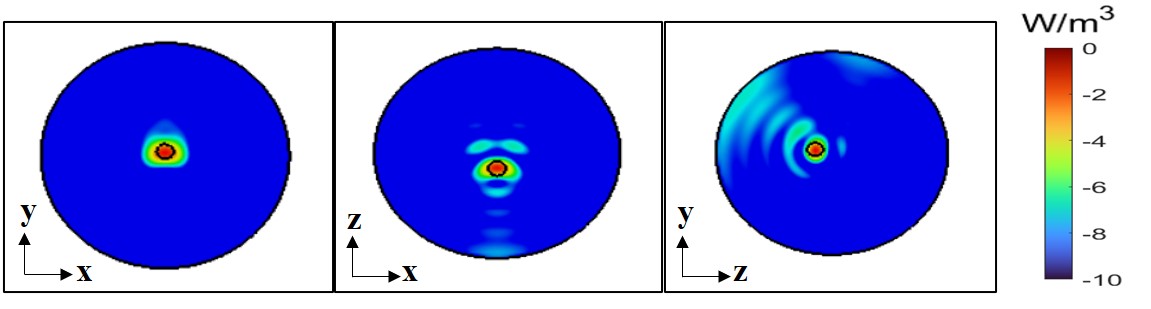} \\
\footnotesize(b) Ideal time-reversal phase sheet ($FoM_Q=0.744$)\\
\end{tabular}
\caption{Normalized power loss density in the spherical head model: cross-sections in x–y (left), x–z (center), and y–z (right) planes at tumor center.}
\label{q_sphere}
\end{figure}

Using CST full-wave simulations with plane-wave excitation, the normalized power loss density was obtained, as shown in Fig.~\ref{q_sphere}. The results demonstrate that the metasurface can focus electromagnetic fields at the tumor location with minimal unwanted hot spots in healthy tissues. The calculated $\mathrm{FoM_Q}$ was below unity, confirming the capability of the system for safe hyperthermia applications.

\subsubsection{Thermal Analysis with the Bio-Heat Equation}
Heat distribution in tissue was analyzed using the Pennes bio-heat equation \cite{pennes}, given in Eq.~(\ref{thermal}). Table IV lists the thermal properties of the brain tissue used in the model.

\begin{equation} \label{thermal}
\rho c \frac{\partial T}{\partial t} = \nabla \cdot (k \nabla T) + \rho Q + \rho S - \rho_b c_b \rho_w (T - T_b)
\end{equation}

\begin{figure}[h!]
\centering
\begin{tabular}{c}
\includegraphics[width=0.48\textwidth, height=1.2in]{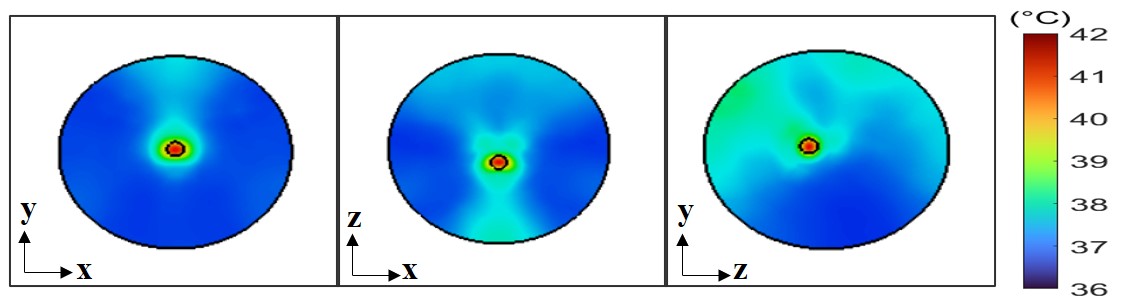} \\
\footnotesize(a) Steady-state temperature distribution in x–y, x–z, and y–z planes centered on the tumor \\
\includegraphics[width=3.2in]{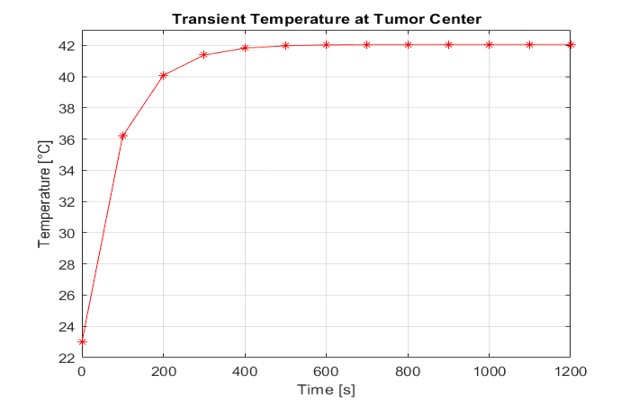} \\
\footnotesize(b) Transient tumor temperature at $(x,y,z)=(0, 2.5, -10)$ mm \\ 
\end{tabular}
\caption{Temperature distribution in the spherical head model after 20 minutes of exposure.}
\label{sphere_thermal}
\end{figure}

The steady-state and transient thermal responses of the tumor and surrounding tissue after 20 minutes of treatment are presented in Fig.~\ref{sphere_thermal}. Results show tumor heating to about $42^{\circ}$C, while surrounding healthy tissue remains near $37^{\circ}$C.

\subsection{System Realization}
\subsubsection{Metasurface Fabrication}
\begin{figure}[h!]
\includegraphics[width=3in]{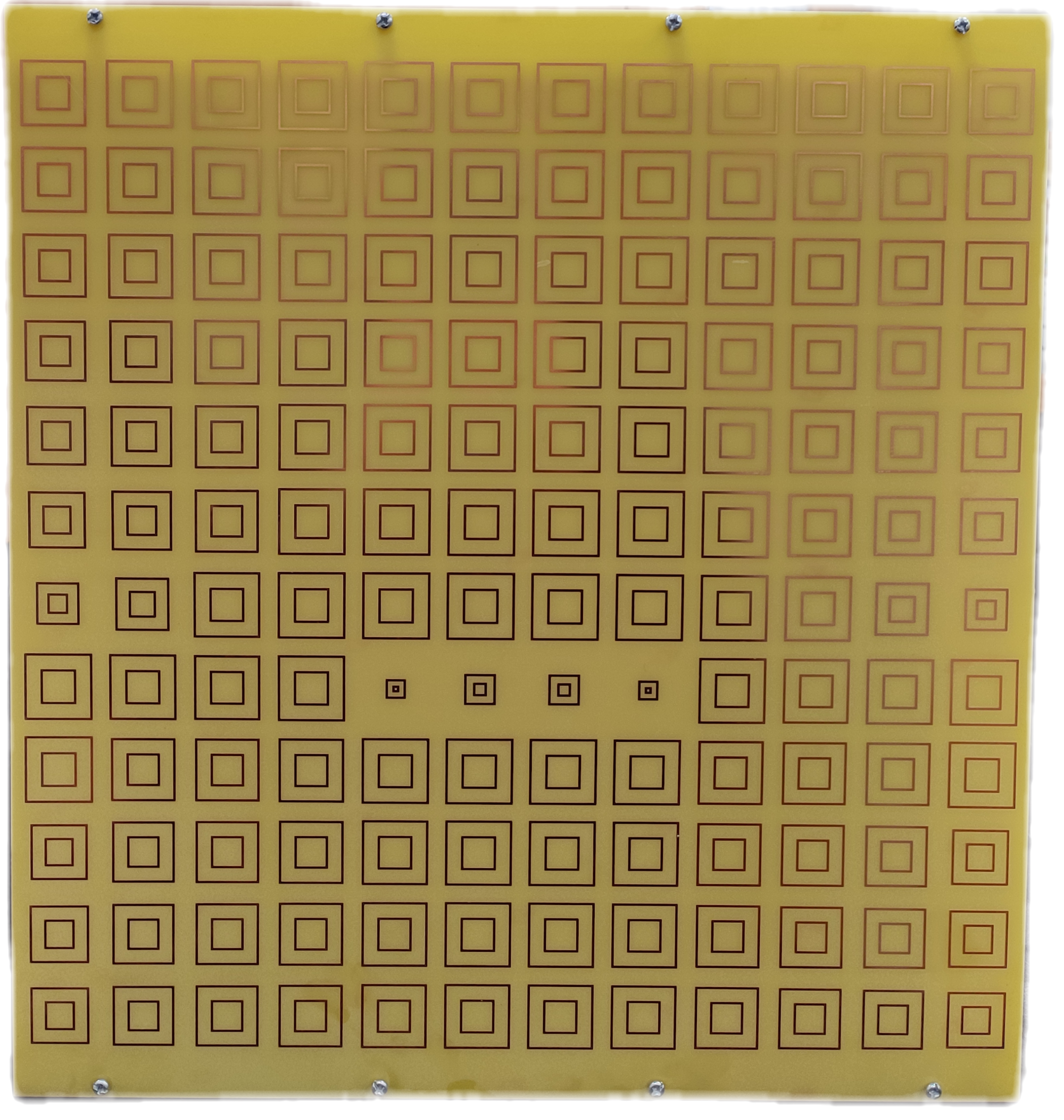}
\centering
\caption{Fabricated metasurface prototype.}
\label{setup}
\end{figure}

The square patch lengths extracted from simulations (Fig.~\ref{sphere_change}(b)) were used to fabricate the metasurface, shown in Fig.~\ref{setup}. Aluminum foil with a thickness of 70 $\mu$m was employed as PEC walls, given its skin depth below 2 $\mu$m at 1.8 GHz.

\subsubsection{Head Phantom Realization}
\begin{figure}[h!]
\begin{tabular}{c}
\includegraphics[width=2.75in,height=1.75in]{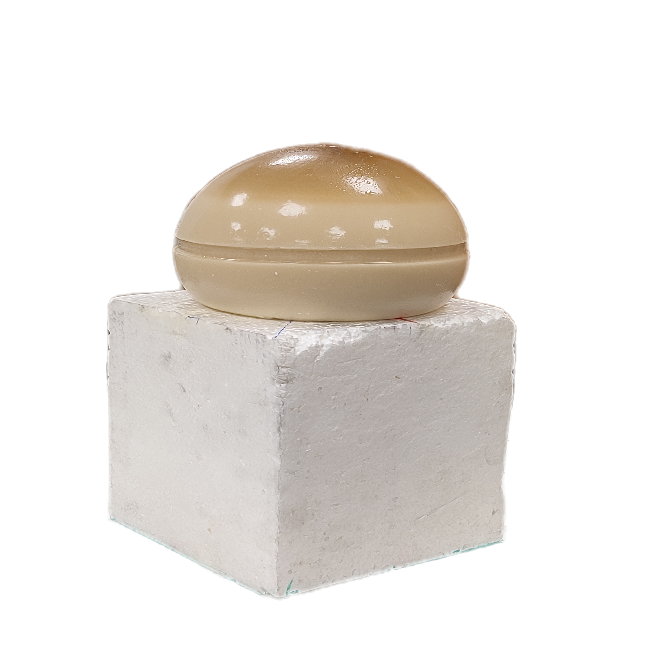} \\
\footnotesize(a) Fabricated spherical head phantom model \\
\includegraphics[width=3.25in,height=2in]{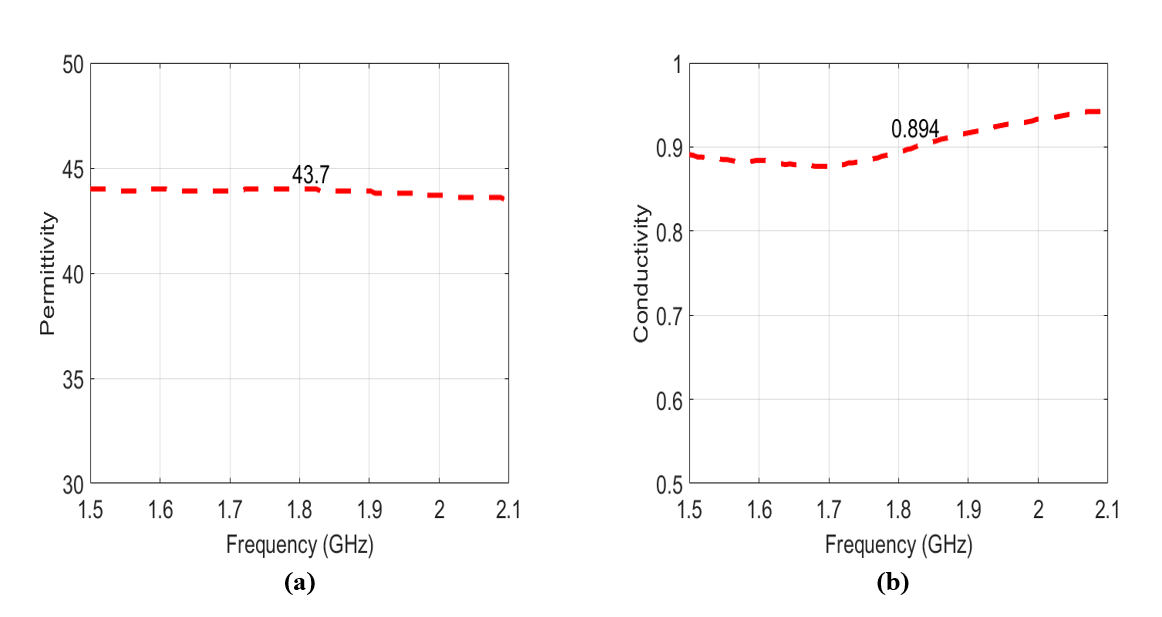} \\
\footnotesize(b) Measured permittivity and conductivity of the phantom mixture \\ 
\end{tabular}
\caption{Realization and electrical characterization of the spherical head phantom.}
\label{phantom_results}
\end{figure}

A thermo-electrical phantom model was developed to replicate both electrical and thermal tissue properties. Using the Bruggeman mixing formula \cite{effective_saviz,effective_medium}, a mixture of distilled water, canola oil, and gelatin powder was designed to achieve permittivity $\varepsilon=43$, conductivity $\sigma=0.9$, and specific heat capacity $C_p=3700$. Table V presents the composition, and Fig.~\ref{phantom_results} illustrates the phantom and its measured electrical properties.

\subsection{Experimental Measurements}
\begin{figure}[h!]
\begin{tabular}{c}
\includegraphics[width=3.25in]{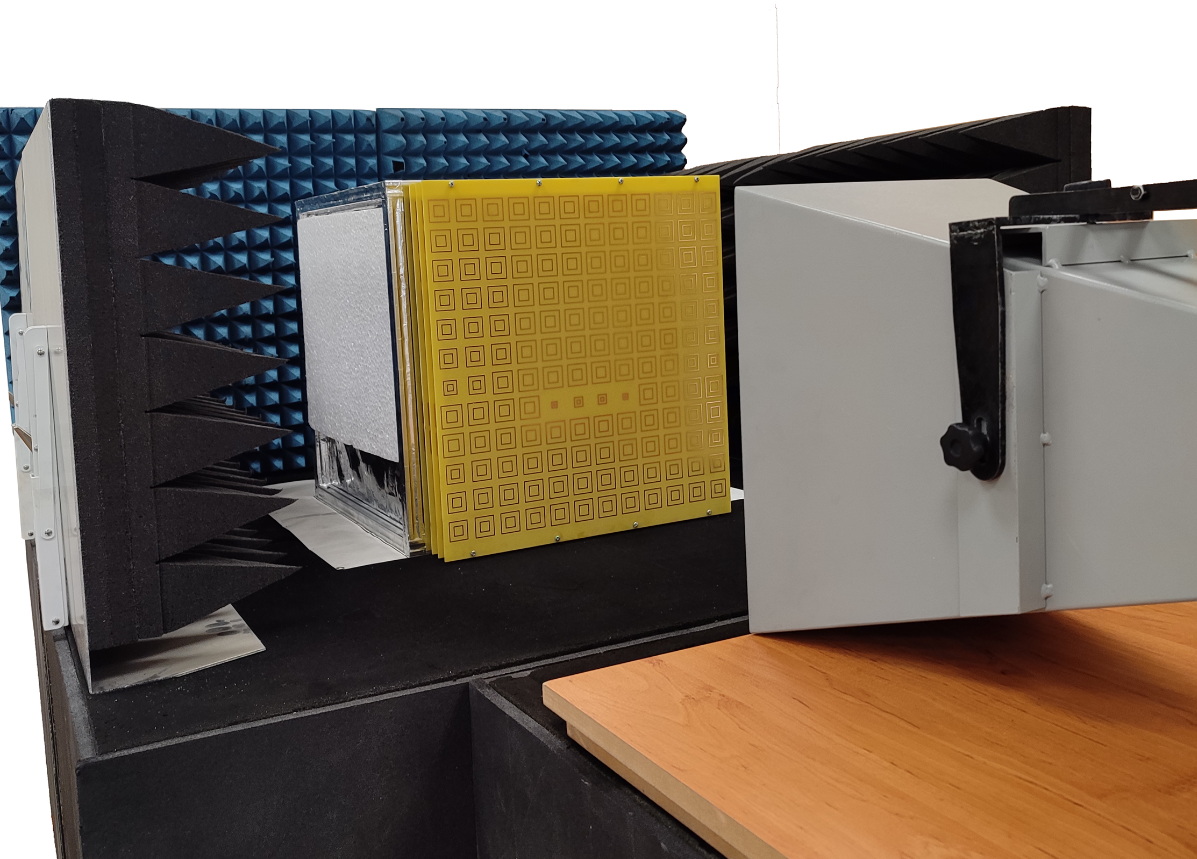} \\
\footnotesize(a) Side view \\
\includegraphics[width=3.5in]{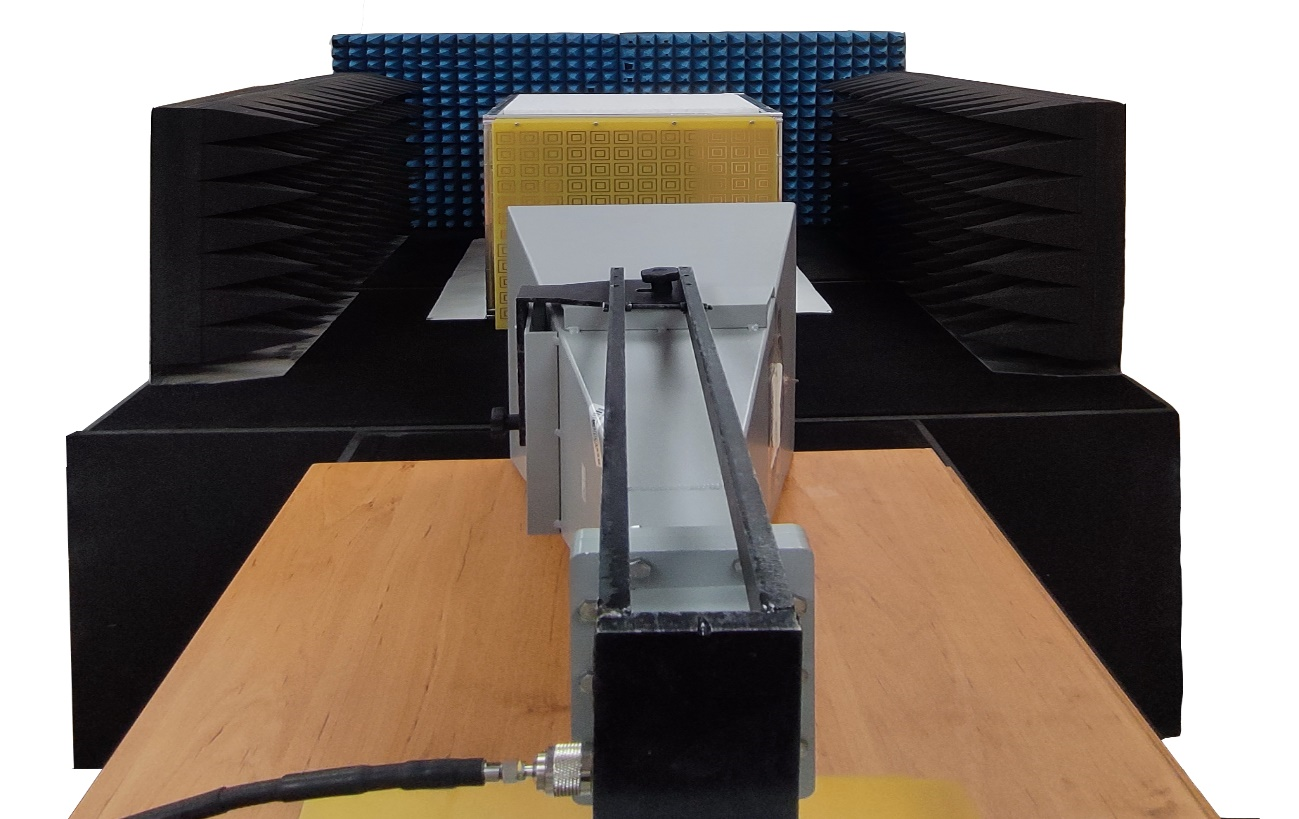} \\
\footnotesize(b) Front view \\ 
\end{tabular}
\caption{Final experimental setup for metasurface-based hyperthermia.}
\label{setup_exp}
\end{figure}

The complete experimental setup is shown in Fig.~\ref{setup_exp}, where the fabricated metasurface, head phantom, and horn antenna excitation are integrated. The horn antenna (ETS 3160-03) operated at 1.7–2.6 GHz, positioned 50 cm from the metasurface based on a trade-off between wavefront planarity and free-space path loss. The combined free-space and substrate losses were estimated at 1.88 dB and 1.58 dB, respectively.

\begin{figure}[h!]
\includegraphics[width=2in]{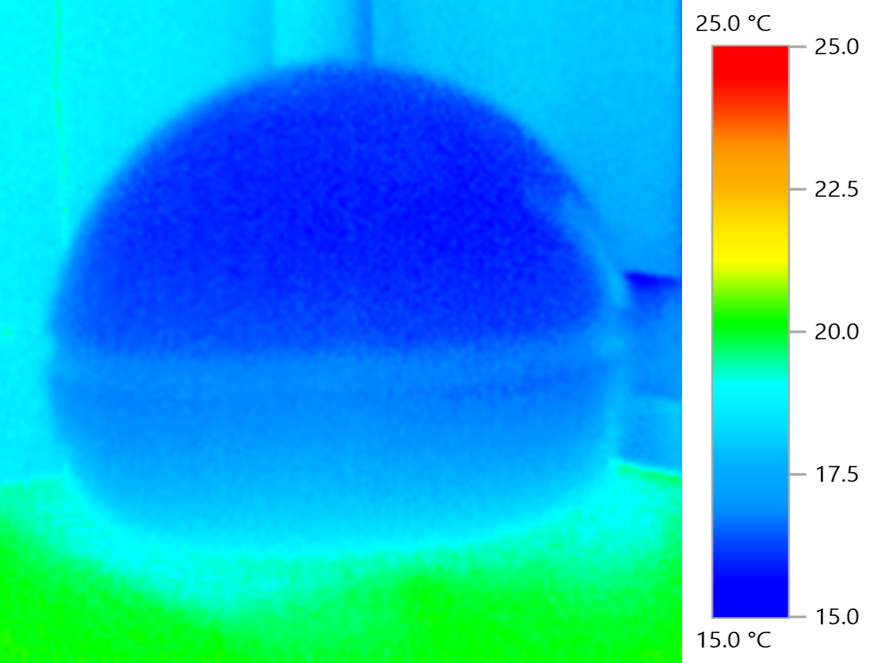}
\centering
\caption{Initial temperature distribution in the phantom prior to heating.}
\label{heat_initial}
\end{figure}

The initial phantom temperature profile, measured using a Testo 872 thermal camera, is shown in Fig.~\ref{heat_initial}, ranging between $16^{\circ}$C and $17^{\circ}$C.

\begin{figure}[h!]
\begin{tabular}{c}
\includegraphics[width=3.35in]{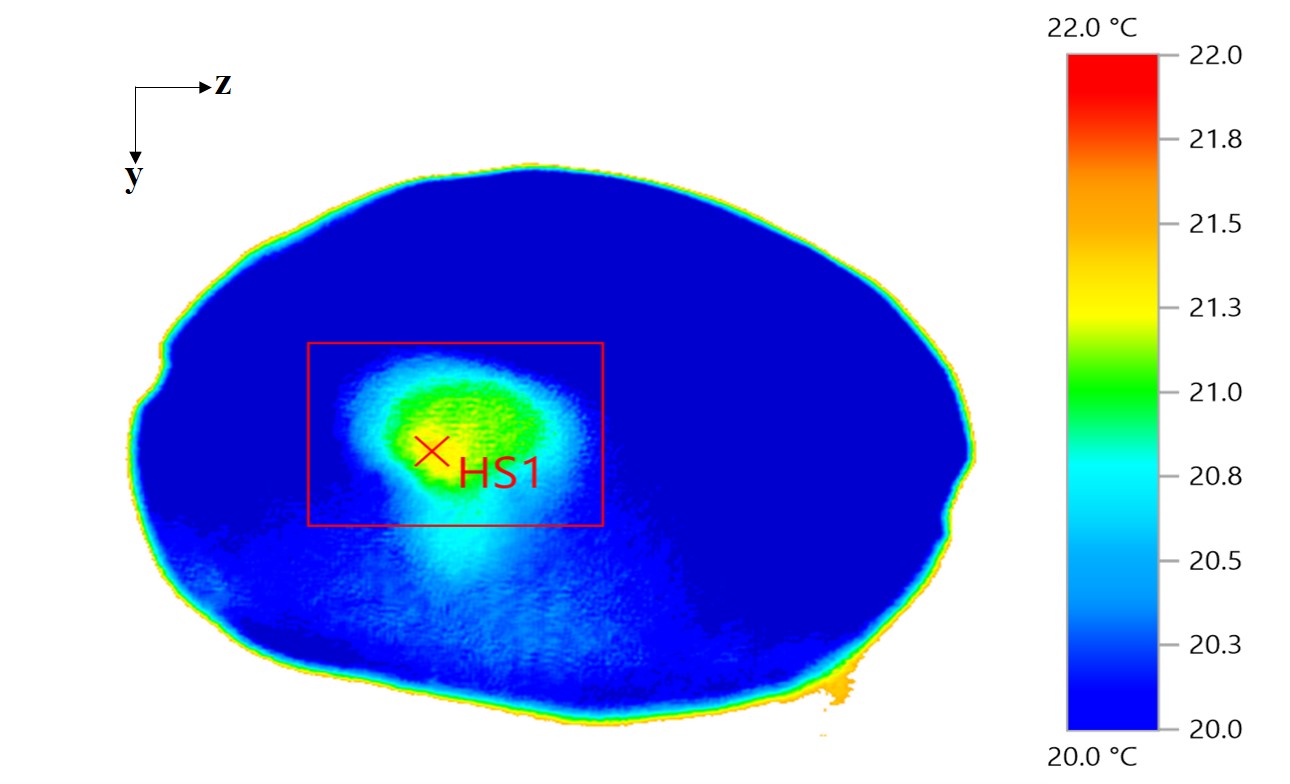} \\
\footnotesize(a) Temperature profile in y–z plane after 20 minutes of exposure \\ 
\includegraphics[width=3in]{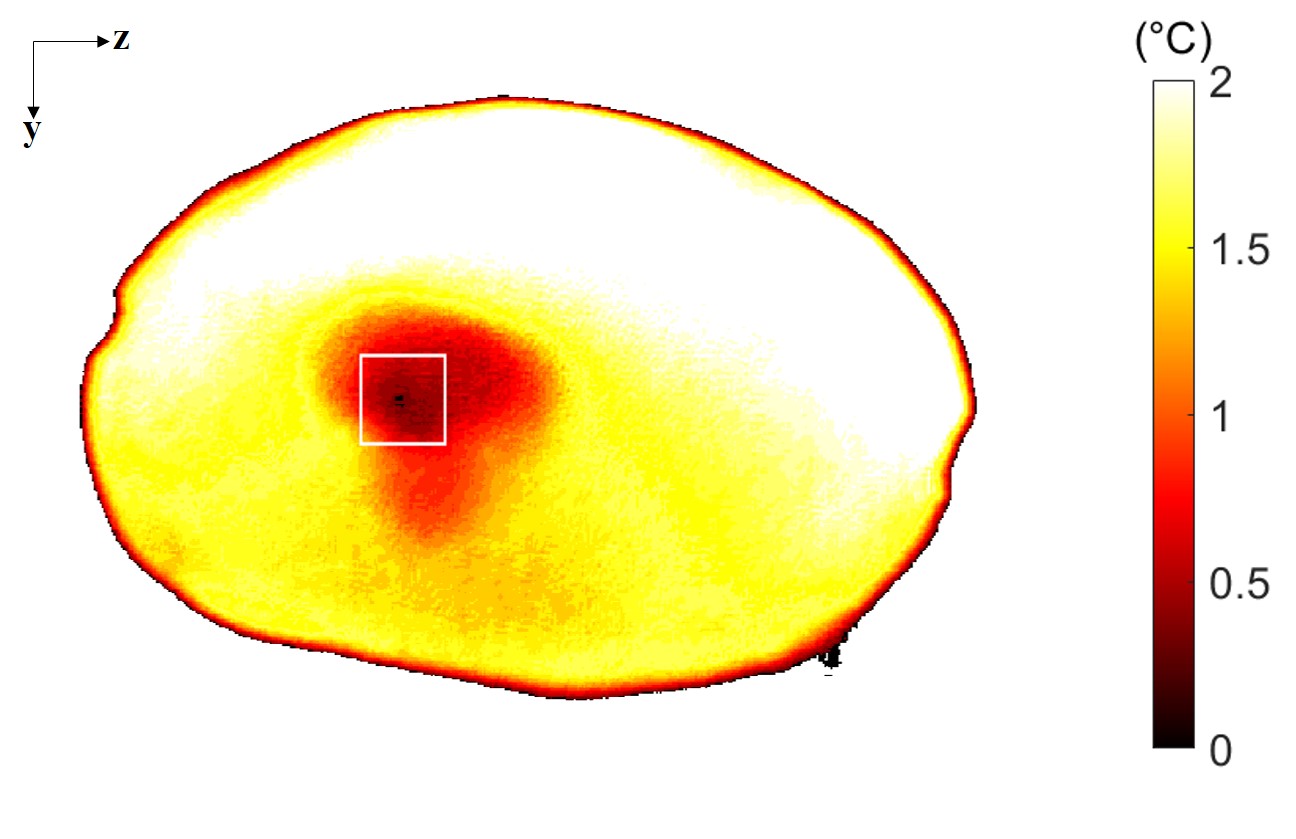} \\
\footnotesize(b) Temperature difference between tumor and surrounding tissues \\ 
\end{tabular}
\caption{Experimental thermal results in the phantom model.}
\label{heat_result}
\end{figure}

After 20 minutes of irradiation, Fig.~\ref{heat_result} shows the resulting thermal profiles. At the tumor center, the temperature increased by $5^{\circ}$C to 21.4°C, while surrounding tissue remained $2^{\circ}$C cooler. The measured hot spot dimensions were 9.29 mm and 9.53 mm along the y and z axes, respectively, closely matching simulation predictions.

Although the phantom’s baseline temperature was lower than natural body temperature (37°C) and biological processes such as blood perfusion were not modeled, the results confirm that the metasurface can focus energy at the tumor site with minimal collateral heating. This validates the feasibility of the proposed metasurface-based hyperthermia system.

\section{Conclusion}
We have proposed a metasurface-based microwave hyperthermia system for deep-seated brain tumors, founded on the time-reversal principle. The unit cell of the transmissive metasurface was designed, and its phase-transfer function was obtained. A numerical investigation was then conducted, introducing a Figure of Merit (FoM) to quantitatively assess system performance. Subsequently, a modified system with a spherical head model was fabricated, and temperature distributions were evaluated. After realizing the phantom head model and extracting its characteristics, experimental results of the temperature profile validated the effectiveness of our systematic metasurface design for microwave hyperthermia. To the best of the authors’ knowledge, this represents the first experimental demonstration of metasurface-enabled time-reversal focusing for hyperthermia in deep-seated brain tumors. This breakthrough opens a pathway toward noninvasive and high-precision thermal therapies, where future developments in conformal metasurfaces~\cite{conformal1,conformal2} may further enhance efficiency and focal resolution.

\ifCLASSOPTIONcaptionsoff
  \newpage
\fi



%

\bibliographystyle{IEEEtran}

\end{document}